# Terahertz detection mechanism and contact capacitance of individual metallic single-walled carbon nanotubes

Joel D. Chudow,[1] Daniel F. Santavicca,[1] Chris B. McKitterick,[2] Daniel E. Prober,[1,2*] and Philip Kim[3]

[1] *Department of Applied Physics, Yale University, New Haven, CT 06511, USA*

[2] *Department of Physics, Yale University, New Haven, CT 06511, USA*

[3] *Department of Physics, Columbia University, New York, NY 10027, USA*

**Abstract**

We characterize the terahertz detection mechanism in antenna-coupled metallic single-walled carbon nanotubes. At low temperature, 4.2 K, a peak in the low-frequency differential resistance is observed at zero bias current due to non-Ohmic contacts. This electrical contact nonlinearity gives rise to the measured terahertz response. By modeling each nanotube contact as a nonlinear resistor in parallel with a capacitor, we determine an upper bound for the value of the contact capacitance that is smaller than previous experimental estimates. The small magnitude of this contact capacitance has favorable implications for the use of carbon nanotubes in high-frequency device applications.

[*] *email: daniel.prober@yale.edu*

Carbon nanotubes (CNTs) have been proposed for a wide range of electronic device applications because of their unique properties.[1] Much work has been done to investigate the high-frequency electrical properties of CNTs in order to assess their potential for use in microwave and terahertz (THz) frequency devices.[2-9] For application as high-frequency detectors, the open issues are the detection mechanisms and the possible limiting effects of the device capacitance. In the present work, we study the effect of the electrical nonlinearity of the contact resistance at low temperature, and confirm this as the detection mechanism at THz frequencies. This enables us to determine that the contact capacitance does not strongly limit the THz performance.

To understand the high-frequency behavior of an individual CNT, one must determine its effective circuit model. The high-frequency circuit model for a single-walled carbon nanotube (SWCNT) was described by Burke.[10] This circuit model is shown in Fig. 1a. The SWCNT is modeled as a transmission line with a kinetic inductance per unit length, $L_K \approx 4$ nH/µm, and a specific capacitance $C_{nt}$ ~50 aF/µm. $C_{nt} = \left( C_{es}^{-1} + C_q^{-1} \right)^{-1}$ is the series combination of the electrostatic capacitance, $C_{es}$, and the quantum (band filling) capacitance, $C_q$. Internal dissipation in the SWCNT can be included as an internal resistance per unit length, $r_{int}$. The characteristic impedance in the limit of small internal resistance is $Z_{char} = \sqrt{L_K/C_{nt}} \approx 10$ kΩ. We include in Fig. 1a the contact resistance $R_c$ and a quantum resistance $R_q/2$ at each end of the SWCNT; $R_q = h/4e^2 \approx 6.5$ kΩ is the two-terminal resistance of the four ballistic quantum channels in parallel, with $h$ Planck's constant and $e$ the electron charge. The contact resistance $R_c$ and its associated parallel contact capacitance $C_c$ arise from the imperfect transparency of the metal/nanotube interface; $R_c$ is zero for a perfect contact.[11] The internal resistance of a SWCNT can be as low as $r_{int}$ =1 kΩ/μm at low temperature.[12]

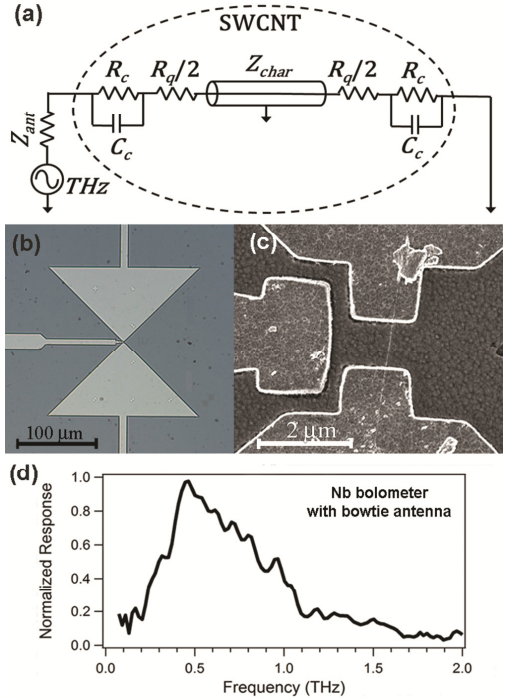

FIG 1. (a) Equivalent circuit model of a SWCNT in an antenna with impedance $Z_{ant}$. (b) Optical and (c) SEM image of sample NT-2 contacted across the gap of a bowtie antenna with a side-gate. (d) Spectral response of the bowtie antenna measured with a Nb bolometer in Fourier transform spectrometer; we find a bandwidth $B$ =0.6 THz peaked at 0.5 THz.

The effective circuit model for an individual SWCNT was studied in previous measurements of the microwave impedance;[13,14] these are challenging measurements because the SWCNT impedance (≳ 10 kΩ) is much greater than the instrument impedance (50 Ω). These studies deduced a lumped-element capacitance between the full SWCNT and the metallic contacts of ~1-10 fF. This was compared to the predicted value of the SWCNT electrostatic capacitance from Ref. 10. However, the model of the electrostatic capacitance as a lumped element in parallel with the contact resistant is incorrect at THz frequency; the model described above, Fig. 1a, should instead be used. Other measurements of $C_c$ include dc measurements of individual semiconducting SWCNTs in the quantum dot regime inferring $C_c \approx 15$ aF,[15] capacitance-voltage measurements of a CNT with a chrome-nanotube Shottky contact inferring $C_c \approx 5$ aF,[16] and microwave rectification measurements of a CNT Schottky diode estimating $C_c$ in the aF range.[17]

We previously reported the detection of rf (≈100 MHz) electromagnetic radiation by an individual SWCNT.[9] The rf detection at bath temperature $T_b = 4.2$ K is due to photon heating of the temperature-dependent resistance (bolometric detection) at higher currents, $|I_{dc}| \geq 0.4$ µA for that sample, denoted NT-1. At lower currents, $|I_{dc}| < 0.4$ µA, detection is due to the nonlinear current-voltage (*I-V*) characteristic of the contact resistance. The heating effect is negligible at these small currents. At $T_b = 77$ K, only bolometric detection was observed. The relative contribution of each mechanism depends on bias current, temperature and frequency. These rf experiments used a SWCNT with very small contact resistance, $R_c \approx 1$ kΩ.

Bundles of SWCNTs, possibly understood with a considerably different circuit model, have also been studied. These exhibit a contact nonlinearity that gives rise to detection at microwave frequencies[5] and for a few devices at THz,[7] but bolometric response appears to dominate for THz detection.[6,7] The bolometric mechanism is believed to dominate due to the contact capacitance effectively short circuiting the nonlinear contact resistance at THz frequencies, but not at microwave frequencies. The inferred total contact capacitance of the bundle is ≈1-10 fF.[6,7] In the present work, we find that the THz detection mechanism of the *individual* SWCNTs we have studied ($R_c > 10$ kΩ) arises from the nonlinear *I-V* curve due to non-Ohmic contacts, and not from bolometric detection. By comparing the magnitude of the measured THz response to that calculated from the measured low frequency *I-V* curve, we are able to place an upper bound on the magnitude of $C_c$ that is smaller than these previous experimental estimates.

Our CNTs are grown using chemical vapor deposition with an Fe nanoparticle or template Co/Mo catalyst that grows predominantly SWCNTs, as determined using atomic force microscopy, with tube diameters < 2 nm. Following growth, the CNTs are located relative to a predefined fiducial grid with a scanning electron microscope (SEM). By means of electron-beam lithography and an electron-beam evaporation lift-off process, palladium electrodes are deposited to contact individual SWCNTs. We report on three individual SWCNTs with a range of resistances (Table I); different dc electrical behavior is typical of various growths. Sample NT-1, studied in past work,[9,12,18] is a $\ell$ = 5 μm long section of an individual SWCNT on an oxidized, degenerately-doped silicon substrate. This arrangement only allowed for dc and rf (100 MHz) testing, as the doped substrate absorbs THz; however, the doped substrate does allow for back-gating to increase the nanotube conductance. Samples NT-2 and NT-3 were grown on high-resistivity silicon substrates with a 500 nm thick oxide (SiO$_2$). High-resistivity silicon ($\rho$ > 5 kΩ−cm) is used because it does not absorb at THz frequencies. The electrodes are deposited in a planar bowtie antenna geometry with a separate side-gate electrode as shown in Fig. 1b and 1c. These show an optical and SEM image, respectively, of sample NT-2 with the planar THz antenna geometry and the SWCNT contacted across the gap. The side-gate is used because the high-resistivity silicon does not allow the substrate to be used as a back-gate. For sample NT-2, the SWCNT bridges the 1 μm gap at the antenna feed, but for sample NT-3 the antenna is slightly misaligned, resulting in a nanotube length $\ell \approx$ 4.5 μm.

Prior to performing measurements of the THz respsonse, we characterized the SWCNT samples at low frequency (≈100 Hz). The differential resistance *dV/dI* for samples NT-2 and NT-3 are presented in Fig. 2 as a function of the dc current and bath temperature $T_b$. Characteristics of sample NT-1 have been previously published[9,12,18] and are described briefly below. Samples

NT-1 and NT-2 displayed a significant gate dependence; data shown for these samples were taken with the device in its high conducting state with a gate voltage of -30 V and -6 V, respectively. The conductance of sample NT-3 displayed only a weak dependence on side-gate voltage, which may result in part from shielding of the gate by the electrodes. All data for sample NT-3 were taken with a side-gate voltage of either 0 or -1 V. At low temperature, all devices show a peak in *dV/dI* at zero bias current. This feature, often referred to as a zero-bias anomaly (ZBA), is related to the imperfect transparency of the contacts. At higher currents *dV/dI* is approximately independent of current for samples NT-2 and NT-3. Sample NT-3 has the largest resistance and the most pronounced ZBA feature, whereas sample NT-2 has more moderate values. Sample NT-1, on the doped Si substrate, showed a small ZBA peak, ≈ 2 kΩ at $T_b$ = 4.2K. Only sample NT-1 has small contact resistance, less than $R_q$, but this sample was not suitable for THz detection, as the doped substrate strongly absorbs THz. Sample NT-1, with ℓ = 5 μm, had total internal dc resistance of $R_{int}$ = 5 kΩ at $T_b$ = 4.2 K; we expect that the other two samples have internal dc resistances that are small compared to their other circuit resistances. We therefore define $R_c$ in Fig. 1a to be $2R_c = dV/dI - R_q - R_{int}$, where $R_{int} = r_{int}\,\ell$.

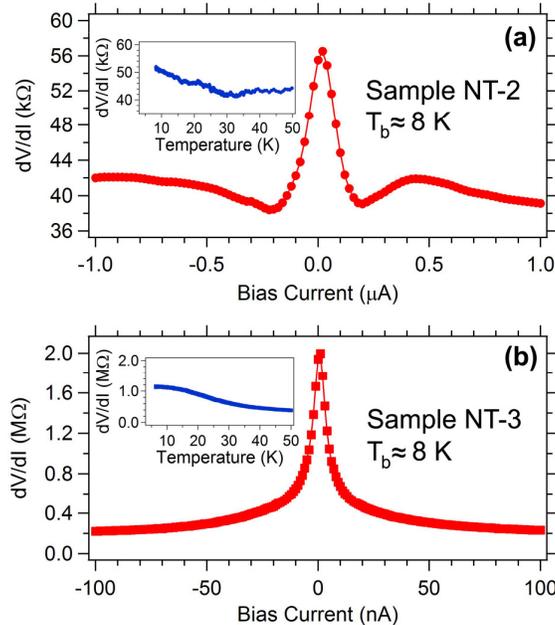

Figure 2. Differential resistance as a function of dc bias current at $T_b$ ≈ 8 K for samples (a) NT-2 and (b) NT-3. The increase in *dV/dI* around zero-bias current is referred to as a zero-bias anomaly (ZBA). Insets: Differential resistance as a function of bath temperature for respective samples measured with (a) 10 nA ac and 50 nA dc bias current and (b) 1 nA ac and 3 nA dc bias current.

In order to couple a THz signal to the SWCNT, we utilize a bowtie antenna with overall dimensions measuring 200 μm by 200 μm (Fig. 1b). This antenna geometry was selected for its simplicity and because it has greater bandwidth than a dipole antenna.[19] Numerical simulations show that the side-gate does not significantly affect the antenna properties. A 6 mm diameter extended hemispherical silicon lens is attached to the back of the substrate for focusing onto the antenna-coupled SWCNT. This configuration takes advantage of the strongly preferential coupling to the antenna through the high dielectric substrate. The device is then mounted on the cold plate of an optical-access liquid-helium cryostat. The THz signal is provided by a silicon carbide globar as a hot blackbody source, coupled through a 6 mm diameter aperture, external to the cryostat with cold infrared low-pass filters that strongly attenuate above 4 THz. Based on the visible spectrum of the globar, we estimate its temperature $T \approx 1300$ K. The signal passes through a THz Fourier-transform spectrometer composed of a Michelson interferometer with a silicon beamsplitter[20] and is focused onto the device lens. The system sits inside a nitrogen drybox to minimize absorption by atmospheric water vapor. A mechanical chopper is positioned in front of the cryostat window, and the device voltage response is measured at the chopping frequency with a lock-in amplifier. The effective temperature of the signal reaching the chopper is taken as $T \approx 650$ K due to the beamsplitter, which only couples half of the globar power to the device. The chopper is at room temperature, but we take its effective temperature to be $T \approx 250$ K to account for partial reflection off the surface of the chopper from the colder cryostat environment. This results in a temperature difference $\Delta T \approx 400$ K.

The THz electrical coupling is understood with the circuit of Fig. 1a, where the antenna, with a low source impedance $Z_{ant} \approx 60\ \Omega$, presents a THz voltage signal to the device. The rms THz voltage difference at the antenna terminals due to the blackbody source chopping is

$$\Delta \langle V_{THz}^2 \rangle = \eta \left( 4 k_B B \Delta T Z_{ant} \right) \quad (1)$$

where $k_B$ is Boltzmann's constant, $B$ is the antenna bandwidth, $\Delta T$ is the source temperature difference, and $\eta$ is the power coupling efficiency of the optical system. Using an impedance-matched antenna-coupled broadband Nb absorber/bolometer, we determine the optical coupling efficiency and the antenna bandwidth. Excluding the interferometer, we found an optical coupling efficiency of $\eta \approx 15\%$.[21] In this measurement, the blackbody source filled the entire field of view of the device lens. This value of $\eta \approx 15\%$ is taken as an upper bound for the coupling efficiency with the interferometer, as any misalignment of the interferometer or mismatch between the interferometer beam pattern and the antenna beam pattern will result in a decreased $\eta$. Equation (1) is in the low-frequency blackbody limit for coupling to a single-mode detector, which is appropriate for our SWCNT at the THz frequency, $f$, since $f \ll k_B T / h \approx 5$ THz for $T = 250$ K. The frequency response of the antenna and the optical system was measured with an Nb bolometer at the bowtie antenna feed instead of the SWNTs. The device voltage response at the chopping frequency is measured as a function of mirror displacement to produce an interferogram, which is then Fourier transformed to determine the spectral response. We find that the frequency response of the antenna plus optical system peaks at 0.5 THz with a bandwidth $B$ of 0.6 THz, as shown in Fig. 1d. Numerical simulations show that the response peak corresponds to the second-order antenna resonance; the first-order resonance is below the low-frequency cut-off of our optical system. Analysis of the SWCNT THz spectral response is in progress and will be presented in a future publication. In the present work, we focus on the dc voltage response to the *total* THz power in the antenna bandwidth and identify the dominant mechanism responsible for the THz-frequency SWCNT device response.

We use an audio frequency lock-in amplifier to measure the dc voltage change due to the modulation of the incident THz power at the chopping frequency ≈100 Hz. At dc and audio frequencies, the sample is biased with a dc current. The high-frequency response due to the non-Ohmic ZBA contact nonlinearity arises from the second-order term of the Taylor series expansion of the *I-V* curve. The dc voltage change when chopping between the two THz blackbody sources is proportional to the second derivative of the *I-V* curve and, assuming the *I-V* curve at THz is the same as at the audio frequency which *dV/dI* is measured, is given by[22]

$$\Delta V_{dc} = (1/2) \Delta \langle I_{THz}^2 \rangle d^2V/dI^2 . \qquad (2)$$

The change in the mean-squared THz current, $\Delta \langle I_{THz}^2 \rangle$, is computed from the equivalent RC

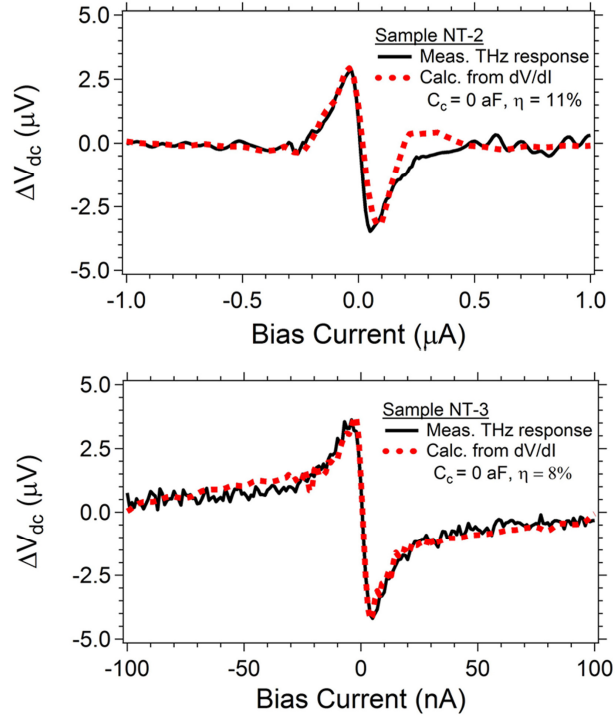

FIG 3. Measured voltage change due to chopped THz source of samples NT-2 and NT-3 plotted against that calculated from Eq. (2) using the measured low-frequency *I-V* curve and the specified optical coupling efficiency η, for $C_c = 0$.

circuit shown in Fig. 1a using $\Delta V_{THz}$ obtained from Eq. (1). We use AWR Microwave Office software to include the transmission line in the calculation. We plot in Fig. 3 the measured and predicted results for $\Delta V_{dc}$, first assuming that $C_c = 0$ and treating η as an adjustable parameter that provides an overall scale factor for optimal curve agreement at low currents. For both samples NT-2 and NT-3, the inferred value of η from this fitting is reasonable with η = 11% and 8%, respectively. The good agreement with theory is evident.

We now consider the possible contribution to the measured response from bolometric detection. The intrinsic voltage responsivity due to bolometric detection, neglecting electrothermal feedback, is given by $S_{bolo} = I_{dc}(dR/dT)/G_{th}$, where $G_{th}$ is the thermal conductance for heat to escape the electron system.[23] From recent work on sample NT-1,[18] we estimate $G_{th}$ ~ 0.1 nW/K per micron for samples NT-2 and NT-3. $dR/dT$ is determined from the measured $R(T)$, Fig. 2. The dc voltage change due to heating of the SWCNT when chopping between the THz blackbody sources is given by

$$\Delta V_{bolo} = I_{dc}(dR/dT)\Delta T_{CNT} = I_{dc}(dR/dT)P/G_{th} \qquad (3)$$

where $\Delta T_{CNT}$ is the SWCNT change in temperature, and $P$ is the power coupled into the SWCNT. For sample NT-1, measured at ≈100 MHz and with its small $R_c < R_q$, the I-V contact nonlinearity response is found in the low current range, but for $|I_{dc}| \geq 0.4$ µA, the bolometric (heating) response is dominant.[9] For samples NT-2 and NT-3, if we assume $\eta = \eta_{max} = 15\%$ we can determine the maximum possible bolometric response. We take $r_{int} = 1$kΩ/µm and we first consider the value of $I_{dc}$ where we observe the greatest THz response. For sample NT-2 at $I_{dc}$ = 50 nA, we compute $\Delta V_{bolo,max}$ ~ 50 nV. Similarly, for sample NT-3 with $I_{dc}$ = 3 nA, we predict $\Delta V_{bolo,max}$ ~ 0.2 nV. These values are much smaller than the measured response as well as the calculated response from the contact nonlinearity, Eq. (2). The bolometric response is expected to increase for larger bias currents, as described by Eq. (3). However, for both samples NT-2 and NT-3, the device noise increases with increasing bias current and we do not observe bolometric detection in the current range presently studied. Additionally, the above analysis assumes that all the absorbed THz power is dissipated in the $R_{int}$ of the SWCNT, with its associated $G_{th}$; however, if the observed *dR/dT* is predominately due to the contacts and not internal to the SWCNT, the power absorption by the comparatively large and thermally anchored contacts, with a much larger $G_{th}$, may result in negligible heating and explain the apparent absence of bolometric response. We conclude that bolometric detection was possible in sample NT-1 due to its low contact resistance, but this mechanism does not contribute significantly for the THz detection observed for higher resistance samples NT-2 and NT-3.

TABLE I. Summary of low temperature sample characteristics. $R_{ZBA}$ is the observed ZBA resistance increase around zero dc bias current. THz response due to the nonlinear *I-V* occurs at low current; bolometric response for sample NT-1 is observed for $|I_{dc}| \geq 0.4$ μA. $C_{c,max}$ is the maximum capacitance determined with a maximum coupling efficiency $\eta_{max} = 15\%$.

| Sample | Length (μm) | Frequency | Observed THz Response | | $R_{ZBA}$ ($T_b$ = 4K) | $C_{c,max}$ ($\eta$ = 15%) |
|---|---|---|---|---|---|---|
| | | | Bolometric | I-V nonlin. | | |
| NT-1 | 5 | 100 MHz | Yes | Yes | 2 kΩ | - |
| NT-2 | 1 | 0.5 THz | No | Yes | 15 kΩ | 70 aF |
| NT-3 | 4.5 | 0.5 THz | No | Yes | 1.8 MΩ | 40 aF |

We now consider the effect of the contact capacitance $C_c$ between the metal electrodes and the SWCNT. Including $C_c$ will reduce the THz voltage and the value of $\Delta V_{dc}$ given by Eq. (2), but does not change the overall shape of $\Delta V_{dc}$ vs. $I_{dc}$ seen in Fig. 3. Consider, for example, sample NT-2, we are able to fit the measured response for $C_c = 0$ with $\eta = 11\%$. If we now consider the limiting case of $\eta_{max} = 15\%$ and $R_{int} = 1$ kΩ, the data are best fit with $C_{c,max} \approx 70$ aF. Similarly, for sample NT-3, $C_{c,max} \approx 40$ aF. These values are the maximum allowed by our model calculation and hence represent an upper bound on $C_c$; the actual value is likely smaller.

A standard figure of merit for THz detectors is the voltage responsivity $S$, defined as the output voltage change divided by the input THz power, $S = \Delta V_{dc}/\Delta P_{THz}$. The internal responsivity $S_{int}$ refers to the THz power coupled into the device, while the external responsivity $S_{ext}$ refers to the available THz power (that which would be coupled into a matched load). We find for samples NT-2 and NT-3 $S_{int} \approx 2$ MV/W and 100 MV/W, respectively, for the maximum nonlinear response ($C_c = 0$). We compute $S_{ext} \approx 10$ kV/W and 15 kV/W for samples NT-2 and NT-3, respectively. We expect for sample NT-1, from the data measured at 100 MHz and assuming that capacitances of the circuit in Fig. 1a would not limit the response, that $S_{ext} \approx 5$ kV/W for the maximum nonlinear response and $\approx 3$ kV/W for the peak bolometric response. These are lower than the values found for samples NT-2 and NT-3 because the nonlinear contact resistance (the ZBA) of sample NT-1 is a much smaller fraction of the total resistance. It should be noted that $S_{ext}$ is a more relevant quantity as compared to $S_{int}$, as this is the responsivity that would be achieved in an actual application.

In summary, we have observed a clear THz response of individual antenna-coupled SWCNTs. We have determined that the mechanism of THz detection is the non-Ohmic contact $I$-$V$ nonlinearity of the two SWCNTs studied, which have resistances of 55 k$\Omega$ and > 1M$\Omega$ at low current. We believe that the lower resistance sample NT-1 would also show THz response if it were on an insulating Si substrate. We have also been able to determine an upper bound for the contact capacitance $C_c$ ~50 aF. This value of $C_c$ is smaller than previous estimates based on microwave impedance measurements of individual SWCNTs[13,14] and measurements of THz detection in CNT bundles.[6,7] This work demonstrates that efficient THz detection utilizing the contact nonlinearity is possible in an individual metallic SWCNT.

*The authors thank M. Brink, B. Connelly, L. Frunzio, M. Purewal, M. Rooks, and M. Takekoshi for assistance and helpful discussions. The work at Yale was supported by NSF under Grant No. DMR-0907082. P.K. acknowledges financial support from NRI supplement through Columbia University NSEC.